\documentclass[proceedings]{JHEP}
\title{\begin{flushright}
       \vspace{-12mm}
        {\normalsize CPHT-PC719.0699}\\[-5mm]
        {\normalsize hep-th/9906108}
\end{flushright}
\vspace {-5mm}
An introduction to perturbative and non-perturbative string
theory\footnote{Lectures given at ``Flavour and Gauge Hierarchies'',
Cargese, France, 20 Jul.-1 Aug. 1998, and at ``Corfu Summer Institute on
Elementary Particle Physics'', Kerkyra, Greece, 6-20 Sep. 1998. }}
\author{Ignatios Antoniadis, Guillaume Ovarlez\\
Centre de Physique Th\'eorique (CNRS UMR 7644), Ecole Polytechnique,
91128 Palaiseau Cedex, France\\}
\abstract{In these lectures we give a brief introduction to perturbative and
non-perturbative string theory. The outline is the following.\\
1.~ Introduction to perturbative string theory\\
1.1 From point particle to extended objects\\
1.2 Free closed and open string spectrum\\
1.3 Compactification on a circle and T-duality\\
1.4 The Superstring: type IIA and IIB\\
1.5 Heterotic string and orbifold compactifications\\
1.6 Type I string theory\\
1.7 Effective field theories\\
References\\
2.~ Introduction to non-perturbative string theory\\
2.1 String solitons\\
2.2 Non-perturbative string dualities\\
2.3 M-theory\\
2.4 Effective field theories and duality tests\\
References}
\begin{document}
\def\be{\begin{equation}}
\def\ee{\end{equation}}
\def\bea{\begin{eqnarray}}
\def\eea{\end{eqnarray}}
\def\noi{\noindent}
\def\nn{\nonumber}
\def\lb{\linebreak}

\section{Introduction to perturbative string theory}

\subsection{From point particle to extended objects}

A p-brane is a p-dimensional spatially extended object, generalizing the
notion of point particles (p=0) to strings (p=1), membranes (p=2), etc.
We will discuss the dynamics of a free p-brane, propagating in $D$ spacetime
dimensions (${\rm p}\le D-1$), using the first quantized approach in close analogy
with the case of point particles. Indeed,
the propagation of a point leads to a world-line. The corresponding action that
describes the dynamics of a free particle is proportional to the length of
this line. The trajectory which minimizes the action in flat 
space is then a straight line.
Similarly, the propagation of a p-brane leads to a (p+1)-dimensional
world-volume. The action describing the dynamics of a free p-brane is then
proportional to the area of the world-volume and its minimization implies that
the classically prefered motion is the one of minimal volume.

More precisely, in order to describe the dynamics of a p-brane in the
embedding $D$-dimen\-sional spacetime, we introduce spacetime
coordinates $X^\mu(\xi_\alpha)$ ($\mu=0,1,..,D-1$) depending on the
world-volume coordinates $\xi_\alpha$ ($\alpha=0,1,..,{\rm p}$).
The Nambu-Gotto action in a flat spacetime is then given by
\be 
S  = -T \int \sqrt{-\mbox{det} h } \,d^{{\rm p}+1} \xi\, ,
\label{ng}
\ee 
where $T$ is the brane tension of dimensionality (mass)$^{{\rm p}+1}$, and
\be 
h_{\alpha \beta} =  \partial_{\alpha} X^\mu \partial_{\beta} X_\mu
\equiv \partial_{\alpha} X^\mu \partial_{\beta} X^\nu \eta_{\mu\nu}
\label{indm}
\ee  
is the induced metric on the brane.

An equivalent but more convenient description is given by the covariant Polyakov
action:
\bea 
S = -{T\over 2} \int d^{{\rm p}+1}\xi\, \sqrt{-\mbox{det} h
}\,\{&&\hskip -0.3cm  h^{\alpha
\beta} \partial_{\alpha} X^\mu \partial_{\beta} X_\mu\nn\\
 &&\hskip -0.3cm -({\rm p}-1)\}\, ,
\label{pol}
\eea 
where the intrinsic metric of the brane world-volume
$h_{\alpha \beta}$ is introduced as an independent variable. Solving the
equations of motion for $X^\mu$ and $h_{\alpha \beta}$ 
\bea
\partial_{\alpha} (\sqrt{-\mbox{det}h}\, h^{\alpha\beta} 
\partial_{\beta} X_\mu ) &=& 0\, ,\label{eqm1}\\
\partial_{\alpha}X^\mu\partial_{\beta}X_\mu -{1\over 2}h_{\alpha \beta} (\partial X)^2
&&\nn\\
+{1\over 2}({\rm p}-1) h_{\alpha \beta} &=& 0\, ,\label{eqm2}
\eea  
with $(\partial X)^2\equiv h^{\alpha\beta} 
\partial_{\alpha} X^\mu \partial_{\beta} X_\mu$, one finds
\be 
h_{\alpha \beta} =  \partial_{\alpha} X^\mu \partial_{\beta}
X_\mu\ ;\qquad {\rm p}\ne 1\, .
\ee  
$h_{\alpha \beta}$ is thus identified with the induced metric (\ref{indm}),
and after substitution into the Polyakov action (\ref{pol})
we obtain back the Nambu-Gotto form (\ref{ng}).

The symmetries of the Polyakov action are (i) global spacetime Lorentz
invariance and (ii) local world-volume reparametrization invariance
under an arbitrary change of coordinates
$\xi^\alpha\rightarrow\xi^{\alpha'}(\xi^\beta)$ with $X^\mu$ and $h_{\alpha
\beta}$ transforming as a (${\rm p}+1$)-dimensional scalar and symmetric tensor,
respectively. The local symmetry needs a gauge fixing: we should
impose ${\rm p}+1$
gauge conditions, so that only the $D-({\rm p}+1)$ transverse to the brane
oscillations are physical.

The equations of motion (\ref{eqm1}, \ref{eqm2}) are generally complicated and
simplify only for the cases of particle (${\rm p}=0$) and string
(${\rm p}=1$).
Indeed, to solve the equations, we apply the gauge-fixing
conditions on the elements of the ${\rm p}\times{\rm p}$ symmetric matrix
$h_{\alpha\beta}$, so that there remain ${1\over 2}{\rm p}({\rm p}+1)$ free components.

For ${\rm p}=0$, one can fix the metric to a constant, $h_{\alpha\beta}=-m^2$.
Furthermore, there is only one world-line coordinate, the time 
$\xi^\alpha=\tau$, and the equation of motion (\ref{eqm1}) becomes
$\ddot{X}^\mu=0$. The general solution is
\be 
X^\mu(\tau)=X^\mu_0+p^\mu\tau\, ,
\ee
which is a straight line. On the other hand, the second equation of motion
(\ref{eqm2}) leads to the constraint
\be
\dot X^2=-m^2\, ,
\ee
which is the on-shell condition $p^2=-m^2$. The quantization can be done 
following the canonical procedure, {\em i.e.} by applying the (equal-time)
commutators
\be
[X^\mu , \dot X^\nu]=i\eta^{\mu\nu}\, ,
\ee
leading to the usual relation $[X^\mu_0,p^\nu]=i\eta^{\mu\nu}$.

For ${\rm p}=1$, we have two world-sheet coordinates $\xi^\alpha=(\tau,\sigma)$, where
$\sigma$ is a parameter along the string. Using the gauge freedom, one can
fix two components of the metric and bring it into the conformally flat form
\be
h_{\alpha\beta}(\sigma,\tau)=e^{\Phi(\sigma,\tau)}\eta_{\alpha\beta}\, ,
\label{conf}
\ee  
where the scale factor $\Phi$ is the only remaining degree of freedom.
However for ${\rm p}=1$, the action (\ref{pol}) has an additional symmetry under
local rescalings of the metric (local conformal invariance), so that $\Phi$
decouples from both the action and the equations of motion (\ref{eqm1},
\ref{eqm2}). As a result, there is an additional gauge freedom and one can
again fix the metric to a constant, $h_{\alpha\beta}=\eta_{\alpha\beta}$, as
in the case ${\rm p}=0$. The equations of motion for $X^\mu$ and constraints then
read:
\bea 
\partial_\alpha\partial^\alpha X^\mu &=& 0\, ,\label{eqm1a}\\
\partial_\alpha X^\mu\partial_\beta X_\mu &=& {1\over 2}
\eta_{\alpha\beta}(\partial X)^2\, .\label{eqm2a}
\eea 

The above equations are further simplified by defining light-cone variables
$\tau,\sigma\rightarrow\sigma_\pm=\tau\pm\sigma$. The equations of motion
(\ref{eqm1a}) then become
\be  
\partial_{+}\partial_{-} X^\mu = 0 \, ,
\label{equmot}
\ee  
and the general solution for $X^\mu(\sigma_\pm)$ is
separated into left- and right-mo\-vers: 
\be  
X^\mu(\sigma_\pm) = X^\mu_L(\sigma_+) + X^\mu_R(\sigma_-)\, .
\label{lr}
\ee 
These functions are subject to the constraints (\ref{eqm2a}):
\be 
(\partial_+X_L)^2=(\partial_-X_R)^2=0\, .
\label{constr}
\ee 
In order to write explicit solutions, we need to impose boundary conditions
that correspond in general to two kind of strings: closed and open.

Closed strings satisfy the periodicity condition
\be 
X^\mu(\tau,\sigma)=X^\mu(\tau,\sigma+2\pi)\, ,
\ee 
leading to the general solution 
\bea \label{csol} 
X^\mu&=& X_0^\mu + P^\mu \tau\\
&+&
{i\over\sqrt 2} \sum_{n \neq 0} {1\over n} \{ a^\mu_n
e^{-in(\tau+\sigma)} +\tilde{a}^\mu_n
e^{-in(\tau-\sigma)}\}\, ,
\nn
\eea  
in mass units of $T={1\over 2\pi}$.\footnote{This convention corresponds to
fixing the Regge slope $\alpha^\prime\equiv{1\over 2\pi T}=1$.} The first two
terms on the r.h.s. describe  the motion of the closed string center of mass,
while the remaining terms in the sum correspond to the string oscillations of
the left- and right-movers which are subject to the constraints (\ref{constr}).
Reality of the coordinates $X^\mu$ imply $(a^{\mu}_n)^{\dagger} =a^\mu_{-n}$
and similar for the right-movers.
Applying the canonical quantization procedure:
\be  
[ X^\mu(\tau,\sigma),\dot X^\nu(\tau,\sigma')] = 
2i\pi \delta(\sigma-\sigma') \eta^{\mu\nu}\, ,
\ee  
one obtains the commutation relations
\bea 
&& [a^\mu_m,a^\nu_n] = [\tilde a^\mu_m,\tilde a^\nu_n]= m
\delta_{m+n,0} \eta^{\mu\nu}\, ,\label{comre}\\ 
&& [\tilde a^\mu_m,a^\nu_n] = 0 \quad\hbox{,}\quad 
[X_0^\mu, P^\nu] = i\eta^{\mu\nu}\, .
\label{comrel}
\eea 

For open strings, the only Lorentz-invariant boundary condition is the
Neumann one:
\be 
\partial_\sigma X^\mu|_{\sigma=0,\pi}=0\, ,
\label{neum}
\ee
implying that the ends of the string propagate with the speed of light.
The general solution (\ref{lr}) becomes in this case
\bea\label{openx}
X^\mu(\sigma,\tau)&=&X_0^\mu + 2P^\mu \tau\\
 &+& i\sqrt{2} \sum_{n \neq
0} {1\over n} a^\mu_n e^{-in\tau} \cos (n\sigma) \, ,\nn
\eea
and can be obtained from the closed string expression (\ref{csol}) by
identifying left- and right-mo\-vers ($a\equiv\tilde a$).\footnote{Note though the factor 2 in $P^\mu$ due to the change in the range of $\sigma$.}

\subsection{Free closed and open string spectrum}

By inspection of the commutation relations\lb (\ref{comre}, \ref{comrel}), one can identify 
$a^\mu_n, {\tilde a}^\mu_n$ with $n>0$ ($a^\mu_{-n}, {\tilde a}^\mu_{-n}$)
with the annihilation (creation) operators. Without loss of generality, let us
first restrict to the holomorphic (left-moving) part. As usual, we can define a
vacuum $|p\!>$ of momentum $p$ annihilated by $a^\mu_n$: 
\be
a^\mu_n |p\!> =0\quad ;\qquad P^\mu |p\!> = p^\mu |p\!>\, .
\ee
The physical states are then created by the action of
$a^\mu_{-n}$'s on this vacuum after imposing the constraints (\ref{constr}).
Expanding the latter into Fourier decomposition, one should require:
\be 
L_m \equiv {1\over 2} \sum_{n \in \rm\bf Z}:a_{m-n} \cdot a_n:=0\ee 
for $m\!>0$, and
\bea\label{Lzero}
L_0&\equiv& {1\over 2\pi}\int^{2\pi}_0 d \sigma({\partial}_+ X)^2\\
&=&{1\over 4}p^2
+{1\over 2}\sum_{n \neq 0}:a_{-n}\cdot a_n:={1\over 4}p^2+N=c\, ,\nn
\eea
where $N$ is the number operator and the constant $c$ appears due to the normal
ordering. These constraints are in fact necessary to eliminate the negative
norm states generated by the action of the time component of creation operators
due to the commutator (\ref{comre}); for instance 
$||\!\,a^\mu_{-m}|p\!>\,\!||=m\eta^{\mu\mu}<0$ for $\mu=0$.

Unitarity is manifest in the light-cone gauge which fixes the residual
symmetry of the covariant gauge fixed equations (\ref{equmot}) and
(\ref{constr}) under $\sigma_+\rightarrow\sigma_+'(\sigma_+)$ and
$\sigma_-\rightarrow\sigma_-'(\sigma_-)$. Defining
\be 
X^\pm=X^0\pm X^{D-1}\, ,
\ee 
this invariance can be used to set:
\be 
X^+=X_0^++p^+\tau\, ,
\ee 
while the constraints can be solved for $X^-$ in terms of the transverse
coordinates $X^T$: 
\be 
\partial_+ X^-={2\over p^+}(\partial_+ X^T)^2\, .
\label{Xminus}
\ee 
Thus, we are left only with the transverse $a^T_{-n}$ that are the independent
physical oscillations.

We can now derive the spectrum. Note that the 0-mode of eq.(\ref{Xminus})
reproduces the global hamiltonian constraint (\ref{Lzero}) that gives the mass
formula:
\be 
-{1\over 4}p^2=N-c\, .
\label{massfb}
\ee 
The first excited state $a^T_{-1}|p\!>$ with $N=1$ is a spacetime vector with
$D-2$ transverse independent components. Therefore, by Lorentz invariance, it
should be massless, implying $c=1$. Furthermore, the quantum algebra of Lorentz
generators -- or equivalently the absence of conformal anomaly -- fixes the
spacetime dimensionality $D=26$. As a result, one obtains:
\be
\begin{array}{lcll} 
N=0 & |p\!> & \mbox{-}{1\over 4}p^2=\mbox{-}1 & \hbox{ tachyon}\\ 
N=1 & a^\mu_{-1}|p\!> &\mbox{-}{1\over 4}p^2=0 & \hbox{ massless vector}\\
N=2 & \left\{\begin{array}{c}a^\mu_{-1}a^\nu_{-1}|p\!>\\  a^\mu_{-2}|p\!>\end{array}\right. & \mbox{-}{1\over 4}p^2=1 &
\hbox{ massive spin 2}\, ,
\end{array}
\ee
and so on. Note that the states of highest spin $J$ at each mass level form a
Regge trajectory: 
\be 
{1\over 4} M^2 = {(J-1)\over\alpha'}\, .
\ee 

The spectrum of a closed string is obtained by the direct product of states
from left- and right-movers with the condition $M_L^2=M_R^2$. At the massless
level, one thus has:
\be  
a^\mu_{-1}\tilde a^\nu_{-1}|p\!>\, ,
\ee 
that can be decomposed into a spin 2 graviton (symmetric traceless part), a
scalar dilaton\lb (trace), and a 2-index antisymmetric tensor (2-form).
The automatic and unavoidable occurrence of the graviton in the massless
spectrum is of course welcome and constitutes a strong motivation for string
theory as quantum theory of gravity.

For open strings, there are no separate left- and right-movers, but one has the
freedom to introduce additional quantum numbers associated to their ends. As a
result the vacuum of oscillators becomes $|p,ij\!>$, where the indices $i,j$
are Chan-Paton charges ``living" at the two ends of the open string. Their
transformation properties define a (non-abelian) gauge group, and the massless
states are now gauge fields:
\be 
a^\nu_{-1}|p,ij\!>\, .
\ee
They can form antisymmetric, symmetric, or complex representations
corresponding to orthogonal, symplectic or unitary gauge groups, respectively.
Obviously, open strings cannot be a theory of gravitation as they do not
contain a massless graviton in the spectrum. However, in the presence of
interactions, closed strings appear by unitarity making string theory as a
candidate for unification of gauge with gravitational forces.

String interactions of splitting and joining can be introduced in analogy with
the first quantized approach of point particles, where the worldline can split
using $n$-point vertices. However, unlike the particle case, the string
interaction vertex is unique modulo world-sheet reparame\-tri\-zations. As a
result, interactions correspond to world-sheets with non-trivial topology and
string perturbation theory becomes a topological expansion in the number of
handles (for oriented closed strings), as well as holes and crosscaps (for
open strings and orientation flips of closed strings, respectively). As we will see in section 1.7, string diagrams are weighted by powers of a coupling constant $\lambda^{2g-2}$, with $g=n+(h+c)/2$ the genus of the surface given in terms of the number of handles $n$, holes $h$ and crosscaps $c$. An important property of string interactions is that (modulo reparametrizations) there is no local notion of interaction point, which makes string perturbation theory free of ultraviolet divergences, that occur in point particles by products of distributions at the same point. Thus, string theory provides a unique mathematical framework of describing non-trivial particle interactions with no ultraviolet divergences.

\subsection{Compactification on a circle and T-duality}

Since the bosonic string lives in 26 dimensions, one has to compactify 22 of
those on some internal manifold of small (presently unobserved) size. The
simplest example of compactification is given by one dimension $X$ on a circle
of radius $R$. $X$ must then satisfy the periodicity condition:\footnote{Here we define $X\equiv X^{25}$.}
\be 
X=X+2\pi R\, ,
\label{per}
\ee  
so that the solution to the equations of motion (\ref{equmot}) is
\bea
 X&=&X_0+p\tau+w\sigma\\
&&+
{i\over{\sqrt 2}} \sum_{n \neq 0} {1\over n} \{ a_n
e^{-in(\tau+\sigma)} +\tilde{a}_n
e^{-in(\tau-\sigma)}\} \nn\\ 
 &=&X_0 + {p+w\over 2}\sigma_++{p-w\over 2}\sigma_-+\hbox{oscillators.}\nn
\eea
Note the appearance of a new term linear in $\sigma$, which is allowed to be
non-vanishing due to the periodicity condition (\ref{per}). Its coefficient 
is\lb quantized in units of $R$, $w=nR$, since for $\sigma\rightarrow\sigma+2\pi$
we must have $X\rightarrow X+2n\pi R$, where $n$ is the (integer) winding
number of the string around the circle. On the other hand, the internal
momentum is also quantized in units of $1/R$, $p=m/R$ for $m\in {\rm\bf Z}$,
which follows from the requirement that plane waves $e^{ipX}$ must be univalued 
around the circle: $X\rightarrow X+2\pi R$.

For closed strings, it is convenient to define left and right momenta: 
\be 
p_{L,R}={m\over R}\pm{n R}\, ,
\ee
in terms of which the mass of physical states becomes (in 25 dimensions): 
\be 
-{1\over 4}p^2={1\over 4} p^2_L+N_L-1={1\over 4} p^2_R+N_R-1\, .
\ee
The above spectrum is invariant under the T-duality symmetry $R\rightarrow 1/R$
with the simultaneous interchange of momenta and windings:
\be 
R\rightarrow{1\over R}\ ,\ m\leftrightarrow n\ \  \hbox{ or }\ \  
X_L\rightarrow X_L \ ,\ X_R\rightarrow -X_R\, .
\label{Tdual}
\ee
It can be shown that T-duality is also an exact symmetry of the string
interactions to all orders of perturbation theory and it is conjectured to hold
at the non-perturbative level, as well. Note that the string coupling 
$\lambda\equiv\lambda_{26}$ should transform, so that the lower dimensional
coupling of the compactified theory $\lambda_{25}={\lambda_{26}/\sqrt R}$
remains inert:
\be
\lambda\rightarrow{\lambda\over R}\, .
\ee

An important consequence of the winding modes in closed strings is the
appearance of enhanced non-abelian gauge symmetries at special values of the
compactification radii. For instance, the generic gauge group in the case of
one dimension is the Kaluza-Klein $U(1)_L\times U(1)_R$:
\bea
&&a^\mu_{-1}|p\!>_L\otimes \tilde a^{25}_{-1}|p\!>_R\ ,\  a^{25}_{-1}|p\!>_L
\otimes \tilde a^\mu_{-1}|p\!>_R \nn\\
&&{\rm with}\ \  p_L^{25}=p_R^{25}=0\, .
\eea
However, for the special value of the radius $R=1$ there is an enhanced gauge
symmetry $SU(2)_L\times SU(2)_R$, due to the appearance of extra massless
states:
\be
\begin{array}{lll}
m=n=\pm 1&\Rightarrow&p_L=\pm 2,p_R=0\ :\\
&&|\pm 2\!>_L\otimes\tilde a^\mu_{-1}|p\!>_R\\
m=-n=\pm 1&\Rightarrow&p_L=0,p_R=\pm 2\ :\\
&&a^\mu_{-1}|p\!>_L\otimes |\pm 2\!>_R
\end{array}
\ee
This property will be generalized later on to non-perturbative states.

For open strings, the Neumann boundary condition (\ref{neum}) imposes the
windings to vanish, $w=0$. Thus, a T-duality is not anymore a symmetry of the
spectrum. Under T-duality, $p\rightarrow w$ ($\sigma\leftrightarrow\tau$), and
the Neumann becomes a Dirichlet boundary condition:
\be
\partial_\tau X|_{\sigma=0,\pi}=0\, ,
\label{diri}
\ee
implying the ends of the string to be fixed at particular points of the
circle. In fact, a coordinate (\ref{openx}) satisfying the Neumann (N)
condition can be written in the form:
\be
X=X_L(\sigma_+)+X_R(\sigma_-)\, ,
\ee
with left- and right-moving oscillators identified, $a_n={\tilde a}_n$. After
the T-duality transformation (\ref{Tdual}), it becomes
\bea
\tilde X&=&X_L-X_R\nn\\ 
&=&{\tilde X}_0 + 2w \sigma\\
&& + i{\sqrt 2} \sum_{n \neq
0} {1\over n} a^\mu_n e^{-in\tau} \sin (n\sigma)\, ,\nn
\eea
which is the general solution of the wave equation (\ref{equmot}) satisfying
the Dirichlet (D) condition (\ref{diri}), implying that the endpoints of the
string are fixed at $\tilde X|_{\sigma=0,\pi}={\tilde X}_0$.

Given the two ends of the open string, one can choose independently the N or D
condition for each endpoint, implying three kinds of boundary conditions: NN,
DD and ND (for which $p=w=0$). The introduction of Dirichlet conditions along
the internal directions leads to the notion of D-branes which are subsurfaces
where open strings can end. A Dp-brane is then defined by imposing Neumann
conditions along its longitudinal directions $X^{\mu=0,..,{\rm p}}$, and
Dirichlet conditions along its transverse ones $X^{I={\rm p}+1,..,25}$. The
endpoints are thus fixed in the $X^I$ transverse space, while they can move
freely in a p+1 dimensional spacetime spanned by the world-vo\-lu\-me of the
p-brane. 

It is now easy to see that performing a T-duality along a compact
transverse (longitudinal) direction, a Dp brane is transformed into a
D(p+1)-brane (D-(p-1)-brane). Moreover, in this picture, the Chan-Paton
multiplicity corresponds to the multiplicity of D-branes. Thus, D-branes on top
of each other, labeled by a Chan-Paton index, give a gauge group enhancement.

\subsection{The Superstring: type IIA and IIB}

Two immediate problems with the bosonic string are the presence of tachyon
(signaling a vacuum instability) and the absence of fermions in the spectrum.
Both problems can be solved in the context of superstring, obtained by adding
fer\-mions on the world-sheet.

The fermionic coordinates are Weyl-Majora\-na two-dimensional (2d) fermions
$\psi^\mu_{L,R}(\sigma,\tau)$,\lb which carry a spacetime index $\mu$. The 2d
Dirac equation implies that the left (right) handed fer-mions depend only on
$\sigma^+$ ($\sigma^-$). As in the case of bosonic coordinates, the timelike
components $\psi^0_{L,R}$ generate extra negative norm states that need a new
local symmetry to be removed. This is indeed the super-reparametrization
invariance, or equivalently local supersymmetry on the world-sheet. In the
superconformal gauge (\ref{conf}), the supersymmetry transformations read
(for the left-movers):
\bea
\delta X^\mu_L&=&-i\epsilon_R\psi^\mu_L\nonumber\\
\delta\psi^\mu_L&=&\epsilon_R\partial_+X^\mu_L
\eea
and emerge from the 2d supercurrent
\be
T_F=\psi^\mu_L\partial_+X_\mu\, .
\label{TF}
\ee
Similar transformations hold for the right movers by exchanging 
$L\leftrightarrow R$ and $+\leftrightarrow -$. The cancellation of conformal
anomalies imply in this case that the superstring must live in $D=10$
spacetime dimensions.

In order to obtain solutions to the equations of motion, we need to discuss
boundary conditions. For both closed and open strings, there are two possible
conditions compatible with spacetime Lorentz invariance:
\be
\psi^\mu(\tau,\sigma)=\pm\psi^\mu(\tau,\sigma+2\pi)\, ,
\ee
corresponding to the Ramond (R) and Neveu-Schwarz (NS) ones. The general
solution can then be expanded as (for instance for the left-movers):
\be
\psi^\mu_L=\sum_r b^\mu_re^{-ir(\tau+\sigma)}\quad ;\quad
(b^\mu_r)^\dagger=b^\mu_{-r}\, ,
\label{psimu}
\ee
where $r$ is half-integer (integer) for NS (R) boundary conditions, and upon
quantization one has the usual canonical anticommutator relations:
\be
\{b^\mu_r,b^\nu_s\}=\eta_{\mu\nu}\delta_{r+s,0}\, .
\label{anti}
\ee
The mass formula (\ref{massfb}) now becomes:
\be 
-{1\over 4}p^2=N+N_\psi-c\, ,
\label{massfs}
\ee 
where $N_\psi={1\over 2} \sum_r r:\!b_r\cdot b_{-r}\!:$ is a sum of half-integer
(integer) frequencies for NS (R) fermions, and $c=1/2$ ($c=0$) in the NS (R)
sector.

As in the bosonic string, unitarity is manifest in the light-cone gauge, where
only the transverse oscillators create physical states. In the superstring,
there are however two sectors in the spectrum: the NS and the R sectors,
corresponding to antiperiodic and periodic world-sheet supercurrent (\ref{TF})
and giving rise to spacetime bosons and fermions, respectively.
In the NS sector, the ground state is still a tachyon $|p\!>$ of momentum $p$
and mass given by $-(1/4) p^2=-1/2$, while at the massless level there is a vector:
\be
|\mu;p\!>=b^\mu_{-1/2}|p\!>\, .
\ee
In the R sector, the fermionic coordinates (\ref{psimu}) have zero-modes which
satisfy the anticommutator relations (\ref{anti}), generating
a ten-dimensional Clifford algebra:
\be
\{b^\mu_0,b^\nu_0\}=\eta^{\mu\nu}\, .
\ee
As a result, the oscillator vacuum forms a representation of this algebra
and corresponds to a spacetime spinor $|p,\alpha\!>$ of dimension $2^{d/2}=32$,
on which $b^\mu_0$ act as the 10d gamma matrices:
\be
i{\sqrt 2}\,b^\mu_0|p,\alpha\!>=\gamma^\mu_{\alpha\beta}|p,\beta\!>\, .
\ee
Moreover, in the absence of oscillators, the constraint corresponding to the
zero-mode of the 2d supercurrent (\ref{TF}) generates the 10d massless Dirac
equation which reduces the dimensionality of the lowest lying state to 16,
corresponding to a massless 10d real (Majorana) fermion.

At this point, the problem of the tachyon remains. However, it turns out that
the above spectrum is not consistent with world-sheet modular invariance, which
guarantees the absence of global anomalies under 2d diffeomorphisms disconnected
from the identity that can be performed in topologically non-trivial surfaces.
Consistency of the theory already at the one-loop level (torus topology) implies
that one should impose the\lb (GSO) projection:
\be
(-)^F=-1\, ,
\ee
where $F$ is the fermion number operator. This projection eliminates the tachyon
from the NS sector, while in the R sector acts as spacetime chirality. In fact,
from the anticommutator
\be
\{(-)^F,b^\mu_0\}=0\, ,
\ee
$(-)^F$ can be identified with the 10d $\gamma^{11}$ matrix, in the absence of
oscillators:
\be
(-)^F=\gamma^{11}(-)^{\sum^\infty_{n=1}b^\mu_{-n}b^\mu_n}\, .
\ee
It follows that at the massless level there is one massless vector and one Weyl
Majorana spinor, having 8 bosonic and 8 fermionic degrees of freedom. This
fermion-bose degeneracy holds to all massive levels and is a consequence of
a resulting spacetime supersymmetry.

We can now discuss the spectrum of closed superstrings. There are two different
theories depending on the relative (spacetime) chirality between left- and
right-movers: the type IIA or type IIB corresponding to opposite or same
chirality, respectively. The massless spectrum is obtained by the tensor
product:
\be
(|\mu\!>,|\alpha\!>_L)\otimes(|\nu\!>,|\beta\!>_{R,L})\, ,
\ee
where $L$, $R$ denotes left, right 10d chiralities, and
$|\beta\!>_R$ ($|\beta\!>_L$) stands for type IIA (type IIB). Decomposing the
spectrum into representations of the 10d little group $SO(8)$, one finds the
bosons
\be
\begin{array}{llll}
|\mu\!>\otimes|\nu\!>&=&1+35_S+28_A&\\
|\alpha\!>\otimes|\beta\!>&\stackrel{\rm IIA}{=}&8_{\rm{(1-form)}}+56_{\rm{(3-form)}} &\;\\
&\stackrel{\rm IIB}{=}&1_{\rm{(0-form)}}+28_{\rm{(2-form)}}+35^+_{\rm{(4-form)}} &\;
\end{array}
\label{nsns}
\ee
and the fermions
\be
\begin{array}{llll}
|\alpha\!>\otimes|\nu\!>&=&8_L+56_L& \\
|\mu\!>\otimes|\beta\!>&\stackrel{\rm IIA}{=}&8_R+56_R &\; \\
&\stackrel{\rm IIB}{=}&8_L+56_L &\;
\end{array}
\label{rr}
\ee
The NS-NS states $1$, $35_S$ and $28_A$ in eq. (\ref{nsns}) denote the trace,
the 2-index symmetric traceless and antisymmetric combinations corresponding to
the dilaton, graviton and antisymmetric tensor, respectively, that are part of
the massless spectrum of any consistent string theory. The above massless
spectra coincide with those of IIA and IIB $N=2$ supergravities in 10
dimensions; note the two gravitini $56$'s in eq. (\ref{rr}) with opposite or
same chirality.

Upon compactification in nine dimensions on a circle of radius $R$, the two
type II theories are equivalent under T-duality:
\be
R\rightarrow{1\over R}\quad :\qquad {\rm IIA}\leftrightarrow {\rm IIB}\, .
\ee
This can be easily seen from the transformation (\ref{Tdual}), whose action
on the fermions is $\psi_L\rightarrow \psi_L$ and $\psi_R\rightarrow -\psi_R$,
in order to preserve the form of the 2d supercurrent (\ref{TF}). As a result, 
$(-)^{F_R}=\gamma^{11}_R\rightarrow-\gamma^{11}_R$,
implying a flip of the fermion chirality in the right-movers.

\subsection{Heterotic string and orbifold compactifications}

The heterotic string is a closed string obtained by the tensor product
of the superstring (for the left-movers) and the bosonic string (for the
right-movers). Since the left-moving coordinates live in 10 dimensions while
the right-moving ones in 26, 16 of the latter are compactified on an internal
momentum lattice. One-loop modular invariance then implies that the
corresponding momenta $p^I_R$ ($I=1,\dots,16$) belong to an even
self-dual lattice (${\vec p}_R\cdot{\vec p}^{\;'}_R\in Z$ and ${\vec p}_R^{\;2}\in 2Z$).
There are only two such lattices in 16 dimensions generated by the roots of
$SO(32)$ and $E_8\times E_8$ groups.

The massless spectrum of the heterotic string is given by the tensor product
\be
(|\mu\!>,|\alpha\!>)\otimes |\nu\!>=1+35_S+28_A+8_L+56_L\, ,
\ee
which forms the particle content of the $N=1$ supergravity multiplet in 10
dimensions, and by
\be
(|\mu\!>,|\alpha\!>_L)\otimes |p^I_R;{\vec p}_R^{\;2}=2\!>\, ,
\ee
which forms a 10d $N=1$ gauge supermultiplet of $SO(32)$ or $E_8\times E_8$.

Upon compactification in nine dimensions on a circle, the two heterotic
theories are equivalent under T-duality:
\be
R\rightarrow{1\over R}\quad :\qquad 
{\rm Het}\ SO(32)\leftrightarrow {\rm Het}\ E_8\times E_8\, .
\label{tdualhet}
\ee
The continuous connection of the two theories can be easily seen using the
freedom of gauge symmetry breaking by turning on Wilson lines (flux lines in
the compact direction). These correspond to the constant values of the
internal (10th) component of the gauge fields $A_9^I$ along the 16 Cartan
generators, and break generically the gauge group to its maximal abelian
subgroup $U(1)^{16}$. One can then show that eq. (\ref{tdualhet}) is valid at
the point with $SO(16)\times SO(16)$ gauge symmetry.

The heterotic string appears to be the only perturbative closed string theory
that can describe our observable world. One of the main problems is however
to get a 4-dimensional superstring which is phenomenologically viable.
In particular, in order to be chiral the spectrum should be at most $N=1$
supersymmetric. For a toroidal compactification, the 10d $N=1$ spectrum is
converted into a non-chiral $N=4$ supersymmetric spectrum in four dimensions.

A solution to this problem is provided by orbifold compactifications. 
These are obtained from toroidal compactifications by identifying\lb points 
under some discrete subgroup of the internal rotations that remains an exact
symmetry of the compactified theory. The resulting compact spaces are not
smooth manifolds because there are singularities associated to the fixed points.
String propagation however is made consistent because modular invariance
requires the presence of a new (twisted) sector corresponding to strings with
center of mass localized at the orbifold fixed points.
In four dimensions, the internal rotations form an $SO(6)\equiv SU(4)$ and the
condition of unbroken $N=1$ supersymmetry amounts to divide the torus 
$T^6$ with a discrete subgroup of $SU(3)$ that leaves one of the four
gravitini invariant.

The simplest orbifold example reduces the number of supersymmetries by 1/2 and
can be studied in 6 dimensions. It is defined by $T^4/\rm {\bf Z}_2$, where 
the ${\rm\bf Z}_2$ inverts the sign of the four internal coordinates 
$X^i\to -X^i$ ($i=1,\dots ,4$) and of their fermionic superpartners. Since in
the heterotic string there are no right-moving superpartners, ${\rm\bf Z}_2$
should also act non-trivially on the gauge degrees of freedom breaking
partly the gauge symmetry together with the $N=4$ supersymmetry. To see the
reduction of supersymmetry in the massless sector, consider the (left-moving)
Ramond vacuum that forms a 10d Weyl Majorana spinor with $(-)^F=-1$. Its
decomposition under $SO(4)\times SO(4)$, with the first factor corresponding to
the 6d little group and the second to the internal rotations, reads
\be
\begin{array}{cc}
\psi^{\mu}&\psi^{i}\nonumber\\
+&-\nonumber\\
-&+\nonumber\end{array}
\ee
where the signs denote the chiralities under the two $SO(4)$. The 
${\rm\bf Z}_2$ orbifold action on this spinor is identical to the chirality
projection in the internal part and thus eliminates half of the gravitini.

The Hilbert space of the theory consists in two sectors:
\begin{itemize}
\item The untwisted sector which is obtained from the Hilbert space of the
toroidal compactification on $T^4$ projected into the ${\rm\bf Z}_2$
invariant states: 
\be
\begin{array}{ll}
|p^i=0\!>_+\ &,\quad
|p^i\!>+|-p^i\!>\\
&\hskip -0.5cm{\rm with\ even\ number\ of\ oscillators}\\
|p^i=0\!>_-\ &,\quad
|p^i\!>-|-p^i\!>\\
&\hskip -0.5cm{\rm with\ odd\ number\ of\ oscillators}
\end{array}
\ee
where $|p^i=0\!>_{\pm}$ denote the ${\rm\bf Z}_2$ even ($+$) and odd ($-$)
states with vanishing internal momentum $p^i$ in which the ${\rm\bf Z}_2$
action is non-trivial.
\item The twisted sector which contains states localized at the $2^4=16$ fixed
points and, thus, are confined to live on 5d subspaces (in the large volume
limit).
\end{itemize}

\subsection{Type I string theory}

Up to this point, we have seen four consistent closed superstring
theories in ten dimensions:\lb type IIA and IIB with two spacetime
supersymmetries, and heterotic $SO(32)$ and $E_8\times E_8$ with one
supersymmetry. Moreover, the two type II theories and the two heterotic ones
are connected by T-duality upon compactification to nine dimensions.
Actually, there is a 5th consistent superstring theory in 10d with $N=1$
supersymmetry, the type I theory of open and closed strings; open strings
provide the gauge sector, while closed strings provide gravity needed for
unitarity.

A consistent algorithm to construct type I theory is to ``orbifolding" type
IIB by the world-sheet involution $\Omega$ that exchanges left- and
right-movers and is a symmetry of the theory:
\be
\Omega\quad : \quad\sigma\rightarrow-\sigma\qquad (L\leftrightarrow R)\, .
\ee
We thus obtain type I theory = IIB/$\Omega$. As in ordinary orbifolds, the
spectrum consists in an untwisted and a twisted sector.

The untwisted sector contains closed strings projected by $\Omega$ 
(unoriented closed strings). This turns to symmetrize the NS-NS sector
and to antisymmetrize the R-R. As a result, the two-index NS-NS antisymmetric
tensor is projected out, together with the R-R scalar and 4-form (see
eq.(\ref{rr})), and we are left with the scalar dilaton scalar $\phi$ and the
symmetric tensor (graviton) $G_{\mu\nu}$ from NS-NS, and a 2-form $B_{\mu\nu}$
from R-R.

The twisted sector corresponds to the ``fixed points"
$X(-\sigma,\tau)=X(\sigma,\tau)$ which are equivalent to the (Neumann) 
boundary conditions for open strings: $\partial_\sigma X|_{\sigma=0,\pi}=0$.
Moreover, the ``fixed-point multiplicity" N corresponds to the Chan-Paton
charges. It is determined by the tadpole cancellation condition, that plays 
the role of modular invariance for open and closed unoriented strings and
guarantees the absence of potential gauge and gravitational anomalies. One
finds N=32 which can also be interpreted as the multiplicity of
D9-branes, leading to an $SO(32)$ gauge group.

Under a T-duality along one compact direction, we get
\be
X=X_L+X_R\rightarrow \tilde X=X_L-X_R
\ee
and the action of $\Omega$($L\leftrightarrow R$) becomes 
\be
\tilde X\rightarrow-\tilde X\, .
\ee
So the effect of a T-duality on $\Omega$ is 
\be
\Omega\rightarrow\Omega{\cal R}\, ,
\ee
where $\cal R$ is defined as ${\cal R}: X\rightarrow-X$.
Therefore, T-duality gives
\bea
\hskip -0.3cm {\rm type\ I=IIB}/\Omega&\rightarrow&{\rm type\ I'=IIA}/\Omega{\cal R}\\
{\rm D9-branes}&\rightarrow& {\rm D8-branes}
\eea

\subsection{Effective field theories}

At low energies, lower than the string scale ${\alpha^\prime}^{-1/2}$, one can
integrate out all massive string modes to obtain an effective field theory
for the massless excitations of the string. There are two ways to obtain
this effective action. Either by computing the string scattering
amplitudes, or by considering string propagation in the presence of
non-trivial background fields. The latter is described by the world-sheet
action:
\bea\label{effec}
S=-{1\over 4\pi}&\int& d^2\xi\{ G_{\mu\nu}(X)\partial_\alpha
X^\mu\partial^\alpha
X^\nu\\
&&+B_{\mu\nu}(X)\epsilon^{\alpha\beta}\partial_\alpha
X^\mu\partial_\beta X^\nu\nn\\
&&-\phi(X){\cal R}^{(2)}+A^a_\mu(X) J^\mu_a+ ...\}\, ,\nn
\eea
where $G_{\mu\nu}$, $B_{\mu\nu}$, $\phi$, $A^a_\mu, \dots$ are backgrounds for
the massless fields (metric, 2-index antisymmetric tensor, dilaton, gauge
fields, etc), and ${\cal R}^{(2)}$ denotes the 2d scalar curvature. Note that
this is the most general non-linear sigma model which is renormalizable in two
dimensions. Conformal invariance implies the vanishing of all beta-func\-tions
which reproduce the spacetime equations of motion for the background fields.

A particular property of string theories, that can be seen from the action
(\ref{effec}), is that the constant dilaton background $e^\phi$ plays the role
of the string coupling. Indeed, a shift $\phi\rightarrow\phi+c$ amounts to
multiply the path integral by a factor
\be
e^{-S}\rightarrow e^{2c(g-1)}e^{-S}\, ,
\ee
where $g$ is the genus of the world-sheet, and we used the Euler integral
\be
{1\over 4\pi}\int d^2\xi{\cal R}^{(2)}=2(g-1)\, .
\ee
It follows that the dilaton shift can be absorbed in a rescaling of the string
coupling $\lambda\to e^c\lambda$. The 10d effective action can therefore by
expanded in powers of $e^\phi$ corresponding to the perturbative topological
string expansion:
\bea
S_{eff}=\int d^{10}X &\{&e^{-2\phi}[{\cal R}^{(10)}+\cdots]\\
&&+e^{-\phi}[\cdots]+1[\cdots]+\cdots \}\, .\nn
\eea
The first term proportional to $e^{-2\phi}$ corresponds to the tree-level
contribution associated to spherical world-sheet topology ($g=0$), the second
term multiplying $e^{-\phi}$ denotes the disk contribution ($g=1/2$), the third
term proportional to the identity corresponds to the one-loop toroidal topology
($g=1$), and so on. Closed oriented string diagrams give rise to even powers
$e^{2(g-1)\phi}$ with $g$ integer, while closed unoriented and open string
diagrams introduce boundaries and crosscaps having $g$ half-integer and can
lead to odd powers of $e^{\phi}$, as well.

\section{Introduction to non-perturbative string theory}

\subsection{String solitons}

The first step towards a non-perturbative understanding of string theory is to
study the analog of field theory solitons. In this section we will establish
that string solitons are p-brane extended objects and study their main
properties.
As point-particles (0-branes) are electric sources for gauge fields (1-forms),
with a coupling $\int A_\mu dx^\mu$, strings (1-branes) are sources for
two-index antisymmetric tensors (2-forms), with a coupling 
$\int B_{\mu\nu} dx^\mu\wedge dx^\nu$ as displayed in eq. (\ref{effec})),
p-branes can be seen as electric sources for $({\rm p}+1)$-form potentials, with a
coupling:
\be
\int A^{({\rm p}+1)}_{\mu_1...\mu_{{\rm
p}+1}}dx^{\mu_1}\wedge...\wedge dx^{\mu_{{\rm p}+1}}\, ,
\ee
with $\mu_i=0,1,\cdots,{\rm p}$. 

Magnetic sources can be understood in a similar way by performing Poincar\'e
duality to the corresponding field strengths. In fact, a $({\rm p}+1)$-form potential
has a $({\rm p}+2)$ field strength $H^{({\rm p}+2)}=dA^{({\rm p}+1)}$. Hodge duality in $D$
dimensions then gives a $D-({\rm p}+2)$ dual form: 
\be
\tilde H^{(D-{\rm p}-2)}=\epsilon H^{({\rm p}+2)}\, ,
\ee
with $\epsilon$ the corresponding Levi-Civita totally antisymmetric
tensor. This dual field strength is now associated a $(D-3-{\rm p})$-form dual
potential $\tilde H^{(D-{\rm p}-2)}=d\tilde A^{(D-{\rm p}-3)}$, which couples to
$(D-4-{\rm p})$-branes playing the role of ``magnetic" sources for the initial
$({\rm p}+1)$-form potential. As a result, a $({\rm p}+1)$-form potential has p-branes as
electric sources and $(D-4-{\rm p})$-branes as magnetic ones.
Furthermore, the analog of Dirac quantization gives the following relation
between the dual charges $\mu_{\rm p}$ of branes:
\be\label{quant}
\mu_{\rm p}\,\mu_{D-4-{\rm p}}=2\pi n\; ,  n\in\rm\bf Z\, .
\ee

For instance, in $D=4$ dimensions, point particles (${\rm p}=0$) electrically charged
are dual to (point-like) magnetic monopoles, while it is also possible
to have dyons carrying simultaneously non-vanishing electric and magnetic
charges. Similarly, one can have dyonic strings (${\rm p}=1$) in $D=6$, dyonic
membranes (${\rm p}=2$) in $D=8$, and dyonic 3-branes in $D=10$.

Let us now consider the spectra of the various string theories:
\begin{itemize}
\item In the heterotic string, there is a 2-form potential $B_{\mu\nu}$. 
Its electric source is the fundamental string, while its
magnetic source is a solitonic NS 5-brane.
\item In type II strings the same result holds for the NS-NS antisymmetric
tensor. In addition, as we have seen in section 1.4 (eq. (\ref{nsns}), there
are R-R p-form potentials arising in the decomposition of two spinors. In
contrast to the NS-NS 2-form, the R-R potentials have no elementary
(perturbative) sources of either electric or magnetic type. This can be
understood for instance from the vanishing of all amplitudes containing R-R
fields at zero momentum since the corresponding string vertices involve
directly their field strengths.

In type IIA, there are all possible even-form R-R field strengths:
0, 2, 4 and their duals 6, 8, 10. They give rise to odd-form potentials
1, 3, 5, 7, 9, having even p-brane sources ${\rm p} = 0, 2, 4, 6, 8$.

In type IIB, there are all possible odd-form R-R field strengths:
1, 3, 5 and their duals 5, 7, 9, the 5-form being self-dual. They give rise
to even-form potentials 0, 2, 4, 6, 8, having odd p-brane sources 
${\rm p} = -1, 1, 3,\lb 5, 7$. Note the $(-1)$-branes that correspond to instantons.

The R-R p-branes are called Dp-branes because they interact through the
emission of open strings with Dirichlet boundary conditions in the transverse
to the world-volu\-me directions.
\item In type I theory, there is no 2-form from the NS-NS sector, because
of the left-right symmetrization of the spectrum. Similarly, because of the
antisymmetrization of the R-R sector, only the 2-form (and its 6-form dual)
potential survive in the spectrum coupled only to D1 and D5-branes.\footnote{In
fact, there are also 32 D9 branes, which is a special case as there is no
field strength for a 10-form potential.} Using now T-duality to type
I$^\prime$, one can actually generate all types of Dp-branes.
\end {itemize}

All the branes we found above are solutions of the supergravity effective
action. They are 1/2 BPS states breaking 1/2 of the supersymmetries, 
and they form short supermultiplets (analog of massless representations). 
Their mass is determined by their charge due to the
supersymmetry algebra, and using the BPS property the mass formula
receives no quantum corrections.

The effective action of a p-brane (neglecting background fields) is:
\bea
S&=&-T_{\rm p}\int_{({\rm p}+1)}\hskip -0.4cm \sqrt{-\det h}\\
&&-\mu_{\rm p}\int A^{({\rm p}+1)}_{\mu_1...\mu_{p+1}}
dx^{\mu_1}\wedge...\wedge dx^{\mu_{p+1}}\, ,\nn
\eea
with $h_{\alpha \beta} =  \partial_{\alpha} X^\mu \partial_{\beta} X_\mu$.
The BPS property fixes its tension $T_{\rm p}$ to be equal to its charge $\mu_{\rm p}$ in
10d supergravity units:
\be
\mu_{\rm p}=\sqrt{2\kappa_{10}^2}T_{\rm p}\qquad\qquad
2\kappa_{10}^2=(2\pi)^7(\alpha^\prime)^4\lambda^2\, .
\label{bps}
\ee
For Dp-branes,
the values of the tension (or the charge) can be extracted from the one-loop
vacuum amplitude of an open string ending on two D-branes, which can also be
seen in the transverse channel as the propagation of a closed string between
the two D-branes. the result is:
\bea\label{tens}
T_{\rm p}&=&{1\over\lambda\,(2\pi)^p(\alpha')^{1+p\over 2}}\\
\mu_{\rm p}&=&\sqrt{2\pi}(4\pi^2\alpha^\prime)^{3-p\over 2}\, ,\nonumber
\eea
which satisfies the Dirac quantization condition (\ref{quant}) with minimal
charge $n=1$. Notice the non-perturbative factor $1/\lambda$ in the expression
of the D-brane tension (\ref{tens}).

Besides the check of the quantization condition (\ref{quant}), one can
derive a recursion relation for the D-brane tensions by use of T-duality
discussed in section 1.6. In fact, a T-duality transformation along a compact
direction longitudinal to the brane, maps a Dp-brane to a D$({\rm p}-1)$. Wrapping
the Dp-brane around a circle of radius $R$, one gets a $({\rm p}-1)$-brane with
tension $2\pi RT_{\rm p}$. Performing now a T-duality along the circle
($R\rightarrow\alpha'/R$, $\lambda\rightarrow\lambda\sqrt{\alpha'}/R$) and
using the non-perturbative dependence $T_{\rm p}\sim 1/\lambda$, one finds the
relation
\be
2\pi RT_{\rm p}\rightarrow 2\pi\sqrt{\alpha'}\,T_{\rm p}\equiv T_{p-1}\, ,
\label{pp1}
\ee
which is satisfied by the general formula (\ref{tens}).

The tension and charge of the fundamental string are:
\be
T_{fund}={1\over 2\pi\alpha'}\qquad
\mu_{fund}=(2\pi)^{5\over 2}\alpha'\,\lambda\, .
\ee
To find the tension and charge of the NS 5-brane, the magnetic dual of the
fundamental string, we can use the quantization condition (\ref{quant}) with
minimal charge $n=1$ and the BPS relation (\ref{bps}) to deduce:
\be
T_{\rm NS\, 5}={1\over\lambda^2(2\pi)^5(\alpha')^3}\qquad
\mu_{\rm NS\, 5}={1\over(2\pi)^{3\over 2}\alpha'\,\lambda}\, .
\label{ns5}
\ee
Notice that the tension of the NS 5-brane is proportional to $1/\lambda^2$, in
contrast to the $1/\lambda$ factor of D-branes, while the tension of the
fundamental string is of course perturbative (of order unity).

Upon compactification to lower dimensions, on finds two important consequences:
\begin{itemize}
\item A p-brane wrapped around a p-cycle of the compact manifold lead to a
non-perturba\-ti\-ve point-like state with
\be
\rm Mass\ = \ Tension \times Area\ of\ the\ cycle,
\ee
generalizing the notion of string winding modes with mass an integer multiple
of\lb $2\pi R/(2\pi\alpha')$. For non-trivial manifolds, if the cycle shrinks to
zero size, one obtains new non-perturba\-ti\-ve massless states that can be charged
under some gauge fields and give rise to enhanced gauge symmetries. In this
way, it is possible to obtain non-pertur\-ba\-tively interesting
non-abelian gauge\lb groups in type II theories.
\item A p-brane with euclidean world-volume\lb wrapped around a $({\rm p}+1)$-cycle
leads to an instanton and thus can generate non-perturba\-tive corrections to
the effective action, proportional to
\be
e^{-S}\, ,\qquad S={\rm Tension \times world\ volume}\, .
\ee
The NS 5-brane generate typical field theory instanton corrections of order 
$e^{-1/\lambda^2}$, following from the expression of its tension (\ref{ns5}).
On the other hand, D-branes have tension $\sim 1/\lambda$ generating
non-perturbative effects of order $e^{-1/\lambda}$. These are typical stringy
in nature and are much stronger than the field theory ones in the weak coupling
limit.
\end {itemize}

\subsection{Non-perturbative string dualities}

In part 1, we have found five consistent superstring theories in ten dimensions.
Two type II with $N=2$ supersymmetry, and two heterotic and the type I with
$N=1$ supersymmetry. The two type II theories, as well as the two heterotic
ones are related by T-duality upon compactification in nine dimensions, leaving
in principle three independent consistent superstring theories. In perturbation
theory, type II theories are phenomenologically uninteresting since they
contain only gravity in ten dimensions and compactification is not sufficient
to produce the rich particle content of the standard model, while type I theory
appears quite complicated. This singled out the heterotic string as the theory
where most of the effort was devoted. The surprising result of string dualities
is that when non-perturbative effects are taken into account, all superstring
theories are equivalent in the sense that they correspond to different
perturbative vacua of the same underlying theory, called M-theory, which
contains also a new vacuum described by the 11d supergravity.

The S-duality inverts the coupling constant $\lambda\rightarrow 1/\lambda$
and exchanges the role of perturbative states with solitons. T-duality
(\ref{Tdual}) is thus an S-duality from the 2d world-sheet point of view, since
the radius $R$ is the coupling constant of the 2d $\sigma$-model and winding
modes correspond to solitons. In ten dimensions, there are two S-duality
conjectures:
\begin{enumerate}
\item Type I - Het SO(32)
\be
\lambda\leftrightarrow{1\over\lambda}\, ,\qquad
B^H_{\mu\nu}\leftrightarrow B^I_{\mu\nu}\, ,
\label{hetI}
\ee
under which we identify
\bea
{\rm D1-string}&\equiv&{\rm Het\ string}\\
{\rm D5-brane}&\equiv&{\rm Het\ NS\, 5-brane}\, .\nonumber
\eea
By identifying the tension of the D1-brane with the one of the heterotic string
$T_1=1/(\lambda_I\,2\pi\alpha_I')\equiv 1/(2\pi\alpha_H')$, we derive the
relation between the type I and heterotic scales:
\be
\alpha_H'=\lambda_I\alpha_I'\, .
\label{hetIa}
\ee
\item Type IIB is self-dual under the S-duality group $\rm SL(2,{\bf Z})_S$,
acting on $\lambda$ complexified with the R-R scalar (see spectrum
(\ref{nsns})):
\be
\lambda\rightarrow{p\lambda-iq\over ir\lambda+s}
\ee
and
\be
\left(
\begin{array}{c}
B^{NS}_{\mu\nu}\\
B^{RR}_{\mu\nu}
\end{array}
\right)=\left(
\begin{array}{cc}
p & q\\
r & s
\end{array}
\right)
\left(
\begin{array}{c}
B^{NS}_{\mu\nu}\\
B^{RR}_{\mu\nu}
\end{array}
\right)
\ee
with the integer parameters satisfying $ps-qr=1$.
For the particular case $p=s=0$ and $q=-r=1$, one finds the simple S-duality
$\lambda\leftrightarrow 1/\lambda$ and
$ B^{NS}\leftrightarrow B^{RR}$, which correspond to the exchange:
\bea
\rm IIB\ string &\leftrightarrow &\rm D1-string\\
\rm NS\ 5-brane &\leftrightarrow & \rm D5-brane\nonumber
\eea
By comparing the fundamental and D1-\lb string tensions, we get again the
transformation $\alpha'\leftrightarrow\lambda\alpha'$.
\end{enumerate}

To summarize up to this point, we have discussed the following relations:
\bea
N_{\rm susy}=1&:& \rm Het_{E_8\times E_8}\stackrel{T}{\longleftrightarrow}
\rm Het_{SO(32)}\stackrel{S}{\longleftrightarrow}\rm type\ I\nn\\
N_{\rm susy}=2&:& \rm IIA\stackrel{T}{\longleftrightarrow}
\rm IIB\stackrel{S}{\longleftrightarrow}IIB
\eea
which lead to two independent theories according to the number of
supersymmetries. The next question is how to relate theories with
different number of space-time supersymmetries. The answer is after 
compactification on different appropriate manifolds leading in lower dimensions
to the same number of supersymmetries.

The first non-trivial example arises in six dimensions. We can relate the
heterotic string compactified on $T^4$ and type IIA on $K3$, which have both
$N=2$ (non-chiral) supersymmetry in $D=6$, by identifying the following branes:
\be
\begin{array}{ll}
\mbox{Het NS 5-brane wrapped around}\ T^4&\\
&\hskip -1cm\equiv\rm IIA\ string\\
\mbox{IIA NS 5-brane wrapped around}\ K3&\\
&\hskip -1cm\equiv\rm Het\ string
\end{array}
\ee
Comparing the branes tensions $T_{fund}\leftrightarrow T_{\rm NS\, 5}V_4$ with
$V_4$ the volume of the 4d compact manifold, we deduce that the two theories
are related by an S-duality in $D=6$:
\be
\lambda_6\leftrightarrow{1\over\lambda_6}\ \ \ \ \rm and\ \ \ \ 
\alpha'\leftrightarrow\lambda^2_6\,\alpha'\, ,
\label{hetII}
\ee
where $\lambda_6=\lambda/V_4$ is the string coupling in six dimensions.

The two theories have indeed the same 6d massless spectrum, which consists of
the $N=2$ supergravity multiplet coupled to $U(1)^{20}$ abelian vector
multiplets containing $4\times 20=80$ scalar moduli. However there is a
potential problem: on the heterotic side, there are special points in the
moduli space with enhanced gauge symmetries, while on the type IIA side
there are no perturbative states charged under the $U(1)$'s because the latter
come from the R-R sector. The missing states appear in fact non-perturbatively
and correspond to D2-branes wrapped around 2-cycles of $K3$. As we discussed in
the previous section, these states become massless when the cycles shrink to
zero-size, giving rise to enhanced gauge symmetries.

\subsection{M-theory}

The connection between the heterotic and type II theories can also be
understood from eleven dimensions, suggesting the existence of some underlying
fundamental theory, called M-theory,\lb whose low-energy limit is $D=11$
supergravity. In this context, the various string dualities follow from 11d
general coordinate invariance.

$D=11$ supergravity is unique and contains the metric $G$, the gravitino and a
3-form potential $A^{(3)}$. Upon dimensional reduction on a circle, it gives
the $D=10$ IIA supergravity with the following field identification:
\be
\begin{array}{cccccc}
\rm 11D&G_{\mu\nu}&A^{(3)}_{\mu\nu
1\!1}&G_{1\!11\!1}&G_{\mu\,1\!1}&A^{(3)}_{\mu\nu\lambda}\\
\rm IIA&G_{\mu\nu}&B_{\mu\nu}&\phi&A_\mu&A^{(3)}_{\mu\nu\lambda}
\end{array}
\label{MII}
\ee
where $B$ is the NS-NS 2-index antisymmetric tensor, $\phi$ the dilaton and
$A^{({\rm p})}$ the R-R p-form potentials. From the identification of the
dilaton with $G_{1\!11\!1}$, it follows that the type IIA string coupling is
given by the radius of the eleventh dimension:
\be
R_{1\!1}=\lambda\sqrt{\alpha'}\, .
\label{R11}
\ee
The corresponding Kaluza-Klein (KK) states with masses $n/R_{1\!1}\sim
1/\lambda$ are non-perturbative states from the type IIA viewpoint and can be
identified with D0-branes. Indeed D0-branes are charged under the R-R gauge
field $A_\mu$ which is precisely the KK $U(1)$ $G_{\mu\,1\!1}$. Moreover,
the D0-brane tension $T_0=1/\lambda\sqrt{\alpha'}=1/R_{1\!1}$ is the same with
the mass of the lightest KK mode having the minimum charge.
KK states become infinitely heavy and decouple in the type IIA weak coupling
limit. On the other hand, when $\lambda\to\infty$, they become light and the
eleventh dimension opens up ($R_{1\!1}\!\rightarrow\!\infty$). As a result, the
strong coupling limit of type IIA string theory is described by the $D=11$
supegravity.

In order to describe the rest of the type IIA spectrum, 11d supergravity should
be implemen\-ted with (BPS) sources for the 3-form potential, in the context of
the underlying M-theory. A membrane (M2-brane) as ``electric" source and a
M5-brane as a ``magnetic" one. Upon compactification on a circle $S^1$ of radius
$R_{1\!1}$, one obtains the following identification:
\bea\label{MDcor}
\mbox{fundamental\ string}&\equiv&\mbox{ M2-brane\ wrapped\ on}\ S^1\nn\\
\mbox{NS\ 5-brane}&\equiv&\mbox{M5-brane}\nn\\
\mbox{D0-brane}&\equiv&\rm KK\\
\mbox{D2-brane}&\equiv&\mbox{M2-brane}\nn\\
\mbox{D4-brane}&\equiv&\mbox{M5-brane\ wrapped\ on}\ S^1\nn\\
\mbox{D6-brane}&\equiv&\mbox{KK\ monopole}\nn
\eea

The tensions of M-theory branes can be computed in the effective supergravity,
although they are essentially determined by dimensional analysis in terms of
the 11d gravitational constant $\kappa_{1\!1}$:
\bea
\label{M2}
T_{M2}&=&\left({2\pi^2\over\kappa_{1\!1}^2}\right)^{1\over 3}\\
T_{M5}&=&{1\over 2\pi}\left({2\pi^2\over\kappa_{1\!1}^2}\right)^{2\over 3}\, .
\label{M5}
\eea
They satisfy the quantization condition (\ref{quant}) for $n=1$, with
the charges equal to the tensions in 11d supergravity units,
$\mu_{\rm p}=\sqrt{2\kappa_{11}^2}T_{\rm p}$ as in eq. (\ref{bps}).
All type IIA brane tensions can be derived from eqs. (\ref{M2}), (\ref{M5}),
using the identification (\ref{MDcor}), in terms of 2 parameters:
$\kappa_{1\!1}$ and $R_{1\!1}$, or equivalently $\alpha'$ and $\lambda$
(determined by the dimensional reduction).
In fact, the quantization condition (\ref{quant}) leaves us with
three independent tensions that are given in terms of two parameters,
so that there is one non-trivial relation among the tensions.
Since this relation can be understood from T-duality (\ref{pp1}) within type
IIA theory, it follows that T-duality is a consequence of 11d reparametrization
in the context of M-theory.

We have seen above that M-theory compactified on a circle $S^1$ describes the
strong coupling limit of type IIA. One can also argue that M-theory
compactified on a line segment $S^1/\rm {\bf Z}_2$ describes the strong
coupling limit of the heterotic string $E_8\times E_8$.
Besides inverting the 11th coordinate $X_{11}\rightarrow-X_{11}$,
${\rm\bf Z}_2$ changes also the sign of the 3-form
$A^{(3)}\rightarrow -A^{(3)}$. As a result, $G_{\mu\,1\!1}$ and
$A^{(3)}_{\mu\nu\lambda}$ are projected out and the ${\rm\bf Z}_2$ invariant
(untwisted) bosonic spectrum is 
\be
G_{\mu\nu}\, ,\quad A^{(3)}_{\mu\nu 1\!1}\equiv B_{\mu\nu}\, ,\quad
G_{1\!11\!1}\equiv\phi\, ,
\ee
which forms the particle content of $N=1$ supergravity in $D=10$.
This truncation of the spectrum introduces a potential gravitational anomaly
at the two (ten-dimensional) ends of the segment,
which is cancelled by introducing twisted states localized at the two endpoints.
They consist of two 9-brane walls with one $E_8$ gauge factor each.
The distance between the two walls is the radius of the 11th dimension which is
related to the heterotic string coupling by the same relation (\ref{R11}) as in
type IIA.

With this result, we completed the discussion on the connection of all
known consistent superstring theories in the context of M-theory.

\subsection{Effective field theories and duality tests}

Here, we will rederive the duality transformations that relate the different
string theories in various dimensions by studying the effective field theories,
and discuss some duality tests.

The 10d effective lagrangians for heterotic and type I theories are:
\bea
\hskip -0.5cm{\cal L}_H&\sim& e^{-2\phi^H_{10}}[{1\over 2}{\cal R}+{1\over 4}F^2+\cdots]\\
\hskip -0.5cm{\cal L}_I&\sim& e^{-2\phi^I_{10}}[{1\over 2}{\cal
R}+\cdots]+e^{-\phi^I_{10}}{1\over 4}F^2\, ,
\eea
where $\phi_D$ is the $D$-dimensional dilaton, such that $e^{\phi_D}=\lambda_D$
is the $D$-dimensional string coupling, and for simplicity you keep only the
gravitational and gauge kinetic terms.
The difference between the two lagrangians comes from the difference in the
topological expansion of the two theories, as discussed in section 1.7: gauge
fields appear on the sphere in closed strings, at the same order with gravity,
while they appear on the disk in open strings.

Compactifying to $D$ dimensions, the lagran\-gian is multiplied by the internal
volume $V_{10-D}$ (in string units)
\be
{\cal L}\rightarrow {\cal L}\,V_{10-D}\, ,
\ee
and we define the $D$-dimensional dilaton
\be
e^{-2\phi_{D}}\equiv e^{-2\phi_{10}}\,V_{10-D}\, .
\ee
In order to normalize the gravitational kinetic terms, we go to Einstein
frame by rescaling the metric
\be
G_{\mu\nu}\rightarrow G_{\mu\nu}\,e^{{4\over D-2}\phi_{D}}\, .
\ee
The gauge kinetic terms take then the form\lb $F^2/4g^2_D$ with
\be
\begin{array}{llll}
{1/ g^2_D}&=&e^{-{4\over D-2}\phi^H_{D}}&\ \ \rm for\ heterotic \\
{1/ g^2_D}&=&e^{{D-6\over
D-2}\phi^I_{D}}\,({V^I_{10-D}})^{1\over 2}&\ \ \rm for\ type\ I
\end{array}
\ee

We can now deduce the duality transformations in various dimensions by comparing
the\lb gauge couplings in both theories:
\begin{itemize}
\item 
In $D=10$ we have 
\be
e^{-{1\over 2}\phi^H_{10}}\leftrightarrow e^{{1\over 2}\phi^I_{10}}\, ,
\ee
which is an S-duality $\lambda\leftrightarrow 1/\lambda$. On the other hand, by
identifying the Newton's constant on both sides:
\be
{1\over\kappa^2_{10}}=e^{-2\phi_{10}}{1\over(\alpha')^4}\, ,
\ee
we get $\alpha'\leftrightarrow\lambda\alpha'$, thus reproducing the
transformations (\ref{hetI}) and (\ref{hetIa}).
\item 
In $D=6$ we have 
\be
V^{1\over 2}_4\leftrightarrow e^{-\phi_6}\, ,
\label{hetI6}
\ee
which is a U-duality $\lambda\leftrightarrow 1/R^2$, with $R$ defined by
$V_4\equiv R^4$.
\item 
Finally, in $D=4$ one obtains a mixing:
\be
\left(
\begin{array}{c}
e^{-2\phi_{4}}\\
V_6
\end{array}
\right)
\leftrightarrow
\left(
\begin{array}{l}
e^{-\phi_{4}}\,V^{1\over 2}_6\\
e^{-3\phi_{4}}\,V^{-{1\over 2}}_6
\end{array}
\right)
\label{hetI4}
\ee
\end{itemize}

For type IIA compactified on $K3$, the effective field theory lagrangian is:
\be
{\cal L}_{I\!I}\sim e^{-2\phi^{I\!I}_{6}}[{1\over 2}{\cal R}+\cdots]
+{1\over 4}F^2\, ,
\ee
where the absence of dilaton dependence in gauge couplings is due to the fact
that gauge fields are R-R states. Compactifying to $D$ dimensions and going to
the Einstein frame as before, we obtain the gauge coupling:
\be
{1\over g^2_D}=e^{2{D-4\over D-2}\phi^{I\!I}_{D}}\,{V^{I\!I}_{6-D}}\, .
\ee
We can now deduce the dualities between heterotic string on $T^4$ and type
IIA on $K3$:
\begin{itemize}
\item 
In $D=6$ we have 
\be
e^{\phi^{I\!I}_6}\leftrightarrow e^{-\phi^H_6}
\ee
which is an S-duality $\lambda_6\leftrightarrow 1/\lambda_6$. Moreover, by
identifying the Newton's constant on both sides:
\be
{1\over\kappa^2_6}=e^{-2\phi_6}{1\over(\alpha')^2}\, ,
\ee
we get $\alpha'\leftrightarrow\lambda^2_6\alpha'$, and reproduce the
transformations (\ref{hetII}).
\item 
In $D=4$ we have
\be
V^{I\!I}_2=e^{-2\phi^H_4}\, ,
\label{hetII4}
\ee
which is a U-duality $\lambda_4\leftrightarrow 1/R$, with $R$ defined by
$V_2\equiv R^2$.
\end{itemize}

Let us now start from 11 dimensions with the supergravity lagrangian
\be
{\cal L}_M\sim {1\over 2}{\cal R}+(dA^{(3)})^2+\cdots
\ee
Compactifying to 10 dimensions, with the identification (\ref{MII})
$A^{(3)}_{\mu\nu 1\!1}\equiv B_{\mu\nu}$ and $G_{1\!1\,1\!1}\equiv R^2_{1\!1}$,
and rescaling the metric $G_{\mu\nu}\rightarrow G_{\mu\nu}/R_{1\!1}$, we obtain
\be
{\cal L}\sim {1\over R^3_{1\!1}}[{1\over 2}{\cal R}+(dB)^2+{1\over 4}F^2]\, .
\ee
Putting back the 11d mass units and defining
$l_{1\!1}^{9/2}\equiv\kappa_{1\!1}$, we find
\bea
\lambda&=&({R_{1\!1}\over l_{1\!1}})^{3\over 2}\nonumber\\
\alpha'&=&{l_{1\!1}^{\,3}\over R_{1\!1}}\, ,
\eea
which reproduce the relation (\ref{R11}).

One consequence of the relations obtained in this section is that
although S-dualities are non-perturbative in higher dimensions, they may become
``perturbative" after compactification which makes possible to perform various
duality tests by comparing appropriate couplings in the effective action. In
addition, non-renormalization theorems due to extended supersymmetry make in
some cases possible to obtain exact results. 

The first non-trivial example
arises in $D=6$, for string vacua with $N=1$ supersymmetry, obtained by
compactification of heterotic or type I theories on $K3$. In this case, there
are two types of (massless) supermultiplets containg scalars: the tensor
multiplet with one scalar, a self-dual 2-form and a Weyl fermion, and the
hypermultiplet, containing 4 scalars and a Weyl spinor. Moreover, for neutral
hypermultiplets, supersymmetry implies that the low-energy (two-derivative)
effective lagrangian is a direct sum of two pieces, one describing the
interactions of tensors and the other describing the interactions of hypers:
\be
{\cal L}^{N=1}_6={\cal L}^{N=1}_{6,\, \rm tensors}+
{\cal L}^{N=1}_{6,\, \rm hypers}\, .
\ee
It turns out that on the heterotic side, the 6d dilaton
$\phi^H_6$ belongs to a tensor multiplet while the compactification volume
$V^H_4$ belongs to a hypermultiplet. The opposite is true on the type I side,
while duality (\ref{hetI6}) maps the dilaton of the one theory to the
compactification volume of the other. Since tensors cannot mix with neutral
hypers, it follows that ${\cal L}^{N=1}_{6,\, \rm tensors}$ can be computed
exactly by a tree-level computation on type I, while 
${\cal L}^{N=1}_{6,\, \rm hypers}$ can be computed exactly by a tree-level
computation on heterotic. Higher order (perturbative or non-perturbative)
cor-\lb rections are forbidden because they depend on the string coupling and
would bring a dilaton dependence and therefore a mixing. The exact expression
should produce the perturbative and non-pertur\-ba\-tive corrections on the other
side, providing an explicit duality test.

A similar but much more non-trivial example arises in $D=4$, for string vacua
with $N=2$ supersymmetry, obtained by compactifying heterotic or type I on
$K3\times T^2$ and type II on Calabi-Yau. In this case, one has vector
multiplets containing a vector, two scalars and a Dirac fermion, and
hypermultiplets. Moreover, for neutral hypermultiplets, supersymmetry implies a
no-mix\-ing at the level of low-energy lagrangian. Now, the heterotic 4d dilaton
belongs to a vector while the size of $T^2$ belongs to a hyper. The opposite
is true on the type II side, while heterotic -- type II duality (\ref{hetII4})
maps the dilaton of the one theory to the $T^2$ size of the other. As a result, 
${\cal L}^{N=2}_{4,\, \rm vectors}$ can be computed exactly by a tree-level
computation on type II side and should produce the perturbative (which stops at
one-loop) and non-perturbative expansion of the heterotic theory. Finally,
on the type I side, one has to define linear combinations and
duality with heterotic (\ref{hetI4}) is more involved. Both vectors and hypers
can receive type I string corrections.

\acknowledgments

\noindent This work was supported in part by the EEC under TMR contract 
ERBFMRX-CT96-0090.

\end{document}